\documentclass[a4paper,priprint]{elsarticle}
\usepackage[a4paper, left=1.5cm, right=1.5cm, top=1.8cm, bottom=1.8cm]{geometry}
\usepackage{graphicx}
\usepackage{amsmath}
\usepackage{float}
\usepackage{booktabs}
\usepackage{xcolor}
\usepackage[colorlinks=true, linkcolor=blue, urlcolor=blue, citecolor=blue]{hyperref}

\begin{document}

\title{Discrimination of \text{neutron}–$\gamma$ in the low energy regime using machine learning for an EJ-276D plastic scintillator}
\author[a,b]{S.~Panda\corref{cor1}}
\ead{srpanda@barc.gov.in}
\author[a]  {P.~K.~Netrakanti}
\ead{pawankn@barc.gov.in}
\cortext[cor1]{Corresponding author.}

\author[a,b]{S.~P.~Behera}
\author[a]  {R.~R.~Sahu}
\author[a,b]{K.~Kumar}
\author[a]{R.~Sehgal}
\author[a,b]{D.~K.~Mishra}
\author[a,b]{and~V.~Jha}
\address[a]{Nuclear Physics Division, Bhabha Atomic Research Centre, Trombay, Mumbai - 400085}
\address[b]{Homi Bhabha National Institute, Anushaktinagar, Mumbai - 400094}

\begin{abstract}
  In this work, we present results for discrimination of neutron and $\gamma$ events using a plastic scintillator detector with pulse shape discrimination capabilities. Machine learning (ML) algorithms are used to improve the discriminatory power between neutron and $\gamma$ events at lower energy ranges which otherwise are not addressed by the conventional pulse shape discrimination techniques. The use of a multilayer perceptron with Bayesian inference (MLPBNN) and support vector machine (SVM) algorithms are studied using the recorded waveforms from the detector. Input variables are constructed for the ML algorithms, which captures the essence of the differences in the head and tail part of the neutron and $\gamma$ waveforms. A new variable, which utilizes the product of kurtosis and variance calculated from the waveform gives better ranking in terms of separation of neutron and $\gamma$ events. The training and the testing of the ML algorithms are done using an AmBe neutron source. In the lower energy region, the results obtained from the ML predictions are compared with the results obtained from a time of flight (ToF) technique to benchmark the overall performance of the ML algorithms. A reasonable agreement is observed between the results obtained from ML algorithm and the ToF experiment in the studied energy range. The MLPBNN gives better discriminatory power for the neutron and $\gamma$ events than the SVM algorithm.
\end{abstract}

\maketitle

\section{Introduction}
Pulse Shape Discrimination (PSD) is a widely used technique for the particle identification in nuclear and particle physics experiments~\cite{nuclPSD,ProsJINST,junoPSD}, as well as in applications such as reactor environment monitoring~\cite{prospNIMBack} and the national security~\cite{homeSec}. Among the various PSD methodologies, the charge comparison method (CCM) is one of the most reliable techniques for categorizing different types of ionizing radiation.

Liquid organic scintillators are well-suited for PSD applications due to their well-defined scintillation time profiles for different particle interactions. The CCM effectively enhances these differences, making it a go-to method for particle identification in particle physics experiments. In particular, the use of this technique in reactor based neutrino detection experiments and rare event decay searches~\cite{DayaBayNIM,PROSPECT2019,ieee}, where precise discrimination of neutron and $\gamma$ signals are required, to enhance the signal to background strength. In a very short baseline reactor based experiment, where both natural radioactivity and neutron-induced backgrounds (including fast and thermal neutrons) are prevalent, CCM provides a means of separating neutron-induced events from the actual signal of interest. Recently, developments in the plastic scintillator domain are being carried out to improvise the PSD properties~\cite{POZZI201319,ZAITSEVA201897}. Easy to handle, large scale production and modularity are some of the advantages of plastic scintillators and can be efficiently used as core detector for the anti-neutrino detection~\cite{SoLiD,PKN2022} and veto systems for muons and neutrons.

Despite its advantages, the performance of CCM degrades significantly at lower energies, posing a major challenge for experiments demanding high background rejection sensitivities.~As an example, the anti-neutrino experiments at very short baseline, few meters from the reactor core, suffers from the fast neutron background inside the reactor complexes~\cite{TEXONO:2018fse,DayaBay:2016ggj}. With optimal shielding configurations, these fast neutron background rates are reduced but can not be completely suppressed~\cite{MULMULE2018104}. The detection of anti-neutrino events, in these reactor based experiments, is generally through the process of inverse beta decay (IBD), which produces a positron and a neutron. The positron, assigned as the prompt signal, deposits its energy instantaneously followed by annihilation with surrounding electrons emitting two $\gamma$-rays. The neutron after thermalization in the detector medium is captured on the capture agent (like H, Li, Gd nuclei). The de-excitation of the capture agent emanates $\gamma$-rays, which forms the delayed signal. On the contrary, the fast neutrons interact with the detector via proton recoil reaction which mimics the prompt event signature and eventually undergoes thermalization and later gets captured by producing a delayed event signature. This can lead to a potential misidentification of a fast neutron event as an anti-neutrino event, even with the optimal shielding configurations~\cite{DayaBay2012,RONI2025}. Apart from the fast neutron background from the reactor,  high energy neutrons, mostly from cosmic or muon induced production, may produce single or multiple neutrons while interacting with the detector material which usually contains carbon~\cite{ROGERS2019162436}. These secondary neutrons in turn may act as a potential source of additional neutron background in the experiments. Therefore, a comprehensive understanding of fast neutron background is vital for accurate event classification to improve the fidelity of neutrino experiments. The PSD technique is also extensively used in low energy nuclear physics experiments, especially in prompt fission neutron multiplicity measurements~\cite{PFNeutrons}, can also benefit from these PSD based plastic scinitillators for neutron and $\gamma$ events. 

In recent years, machine learning (ML) techniques are increasingly being utilized for particle identification tasks. Supervised learning algorithms such as boosted decision tree (BDT), support vector machine (SVM), convolutional neural network (CNN), artificial neural network (ANN) have shown assuring results in improving discrimination of neutron-$\gamma$ events. The BDT models have shown a low-resource, light-weight algorithms that can be included in online data taking using eFPGA~\cite{Johnson_2024}. SVMs achieve neutron–$\gamma$ separation by finding an optimal hyperplane that best divides the data even for non-linear datasets~\cite{YU201580svm,ARAHMANE2020106958svm}. CNN models are used in various types of organic and inorganic scintillator detectors for efficient neutron detection~\cite{YOON20233925cnn,10672537cnn,Yu:2025lxz,Shen:2025vyt}. Various ANN models are employed with different labeling strategies to capture the complex non-linear behavior of the input dataset, enhancing neutron–$\gamma$ discrimination performance~\cite{HACHEM20234057ann,ZHANG2024111179bpann}. In contrast, unsupervised learning algorithms~\cite{ZHAO2025107483} have also shown promising results for the same task. In this study, supervised ML algorithms are explored to address the inefficiency of PSD CCM in the lower energy regime. Two approaches are investigated, a Multi-Layer Perceptron Bayesian Neural Network (MLPBNN) and a Support Vector Machine (SVM). These algorithms are trained and evaluated using data from a plastic scintillator detector, with PSD capabilities, to assess their ability to distinguish between particle types more effectively at lower energies.
 To check the effectiveness of the considered ML algorithms, we have performed a time-of-flight experiment. The neutron and $\gamma$ events are separated in the time domain and compared with the ML-based discrimination results.

\section{Experimental Setup}
\begin{figure}[H]
  \centering
  \includegraphics[,width=0.7\textwidth]{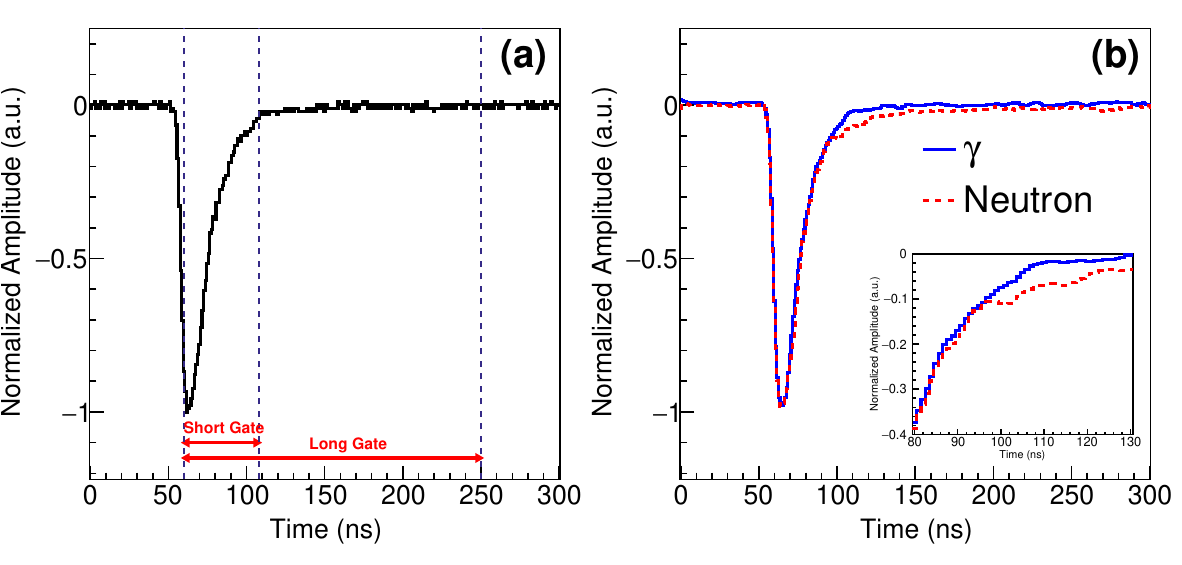}
  \caption{(a) Normalized waveform from a EJ276D PS detector, (b) neutron (red dashed line), and $\gamma$ (blue solid line) waveforms.}
  \label{Waves}
\end{figure}

A Plastic Scintillator (PS), EJ276D with PSD property from Eljen technology~\cite{eljen276d}, was used to discriminate the signals from neutron and $\gamma$ events. The scintillator is of dimension $\phi$51$\times$51 mm and has a maximum emission wavelength of 425 nm. The detector was optically coupled with a $\phi$51 mm diameter Hamamatsu R7724, 10 stage photo-multiplier tube (PMT) of borosilicate glass window, which was operated at the negative bias voltage of 1200 V. The PSD ability of this particular PS were studied before and a decent neutron and $\gamma$ discrimination was obtained for higher energies~\cite{Pant2024,Pagano2024}. 
The data acquisition system (DAQ), used to measure the anode signal from the PS, consists of a 8-channel CAEN V1751 digitizer~\cite{CAEN2024}. On-board digitization of the anode signal from the PMT was done using a leading edge discrimination technique. Triggered data was read by a Flash ADC of 10-bit resolution at 1 GSPS sampling rate. Raw waveforms were recorded and transferred to the computer via optical link for the detailed offline data analysis.
Figure~\ref{Waves}(a) shows a typical baseline subtracted waveform, recorded with PS, measured using an $^{241}\mathrm{Am} ^{9}\mathrm{Be}$ neutron source. The rise time of the scintillation signal is around 2 ns and the waveform acquisition window is kept at 300 ns to capture the entire charge deposition from the ionizing particle. For the determination of the PSD, an integrated charge from short gate (SG) and the long gate (LG) are used which is described in detail later in the text. Normalized neutron and $\gamma$ waveforms, shown in Fig.~\ref{Waves}(b), where the differences in the charge distribution between the two signals is observed at the tail region beyond 100 ns. The inset in Fig.~\ref{Waves} (b) shows the zoomed version of the waveform from neutron and $\gamma$ event in the time window of 80 to 130 ns, to emphasis the differences in the waveforms for the PSD variable calculation.

\begin{figure}[H]
  \centering
  \includegraphics[, width=1.0\textwidth]{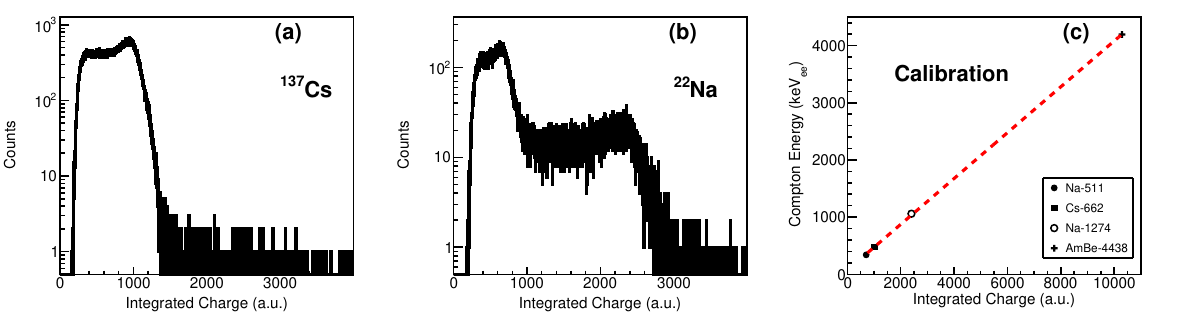}
  \caption{Integrated charge distribution of (a) ${}^{137}\textbf{Cs}$, (b) ${}^{22}\textbf{Na}$ radioactive source and (c) calibration obtained from the measured Compton edge energy deposition in the detector.}
  \label{sourceADC}
\end{figure}

The radioactive $\gamma$ sources were used to study the response of the PS. The integrated charge spectra from the $^{137}\mathrm{Cs}$ and $^{22} \mathrm{Na}$ radioactive sources are shown in the Fig.~\ref{sourceADC} (a) and (b), respectively. The prominent Compton edges can be seen indicating the energy deposition of the $\gamma$-rays in the PS. The PS detector was calibrated using the $\gamma$s  from $^{137}\mathrm{Cs}$ and $^{22} \mathrm{Na}$ sources along with the 4.4 MeV $\gamma$ from an AmBe neutron source, which is shown in the Fig~\ref{sourceADC}(c).~The calibrated energy represents the electron equivalent energy deposited ($\mathrm{E_{dep}}$) by the interacting particle, in units of $\mathrm{keV_{ee}}$, in the detector.
The lowest measured integrated charge value used in our analysis is 200 which corresponds to a $\mathrm{E_{dep}}$ of 151 $\mathrm{keV_{ee}}$. For the analysis of the waveform and calculation of variables for ML algorithm, we have only used the integrated charge values.

The neutron and $\gamma$ data are collected using an AmBe neutron source, kept at a distance of 5cm from the PS detector. No shielding or moderator is used between source and the PS. The neutrons emanating from the AmBe source range from 1 MeV to 12 MeV, peaked around 3-4 MeV. The neutron deposits energy in the PS by proton recoil and the energy deposition is quenched due to the electron density in the PS~\cite{Reichhart:2011gz}. 
On the other hand, a 4.4 MeV $\gamma$ is emitted from the AmBe source produces the Compton scattered electron in the PS. The energy deposition, represented in the waveform, of the proton recoil and the Compton scattered electron is measured from the anode signal of the PMT and digitized using a waveform digitizer. 

\section{Waveform Analysis}
The recorded raw waveform for each event is stored and baseline subtraction is done offline mode. The average value of the charge for the first 46 ns time window is used to calculate the baseline and is subsequently subtracted from the recorded waveform to obtain the final baseline subtracted result, as shown in Fig.~\ref{Waves}(a).

Organic scintillators exhibit fast and slow components due to their decay modes. All charged particles interaction with the scintillator medium produces prompt fluorescence by exiting the singlet states. This process results in the decay time, in the order of a few nanoseconds, called a fast-time component. On the other hand, heavy charge particles, like proton and alpha particles, create higher ionization density of excited molecules in the scintillator medium as compared to electrons. This in turn, increases the probability of excitation of the triplet states leading to the delayed photon emission, called the slow component. The different ratios of fast and slow signals coming from various particles cause the pulses to have different shapes and enables the basis of PSD. Plastic scintillators have a very fast rise time of the $\sim$2 ns and decay time around $\sim$5 ns~\cite{knoll2010radiation}.

The interaction of $\gamma$ with organic scintillator is mediated by electromagnetic interaction, hence, having a fast rise and decay time. For the fast neutrons the dominant mechanism of energy transfer is through proton recoils, producing higher ionization density than electrons from $\gamma$-ray interactions. Due to the heavier mass of the proton, it loses energy slowly in the detector medium in comparison to electron which is generated by Compton interaction of $\gamma$-rays. Also the scintillation response time is slow for the protons due to the quenching mechanism~\cite{Awe_2021}. This results in a comparative longer tail part for the neutron waveforms. This difference in the neutron and $\gamma$ waveform from the EJ-276D PS is shown in Fig.~\ref{Waves} (b).
The traditional CCM exploits the timing nature of different particles. Considering two distinct time windows SG and LG, charge is integrated to obtain $Q_S$  and $Q_L$, respectively. The ratio of $Q_S$ to $Q_L$, hence enables the pulse shape discrimination capability of the detector. The PSD variable, $PSD = 1-\frac{Q_S}{Q_L}$, is calculated from each waveform and studied as a function of the total charge deposition in scintillator for the discrimination of neutron and $\gamma$ signals. Figure of Merit (F.o.M), described as $\mathrm{F.o.M = \frac{|\mu_\gamma - \mu_n|}{FWHM_\gamma+FWHM_n}}$, is often used to estimate the degree of discrimination between the different particles. The $\mathrm{\mu_\gamma}$ and $\mathrm{\mu_n}$ represent the mean values of the neutron and $\gamma$ PSD distribution and $\mathrm{FWHM_\gamma}$ and $\mathrm{FWHM_n}$ represents the full width at half maximum of these distributions, respectively.
To obtain a reasonable F.o.M, various combinations of $Q_S$ and $Q_L$ gate time windows are used in an iterative manner in the analysis.

The optimized SG and LG corresponds to a time window of 60-108 ns and 60-250 ns, respectively. In the integrated charge range 1500-4000, a F.o.M of 1.21 is achieved, where the PSD discrimination still suffers from the non uniform energy loss by the proton recoils from neutron interactions in the detector. At low deposited energies, the number of detected photons is small, and statistical fluctuations in the fast and slow components reduce the neutron-$\gamma$ discrimination. In contrast, the F.o.M in the integrated charge range 2000-4000 improves to 1.35, implying when the energy of the particle increases, more photons are generated, which reduces the  statistical fluctuations and both the fast component and slow component increase proportionally assuming the ionization density remains the same in both cases. Due to this, at higher charge region, the PSD is constant as a function of deposition energy in the measured range~\cite{sym15010011}. This shows the intrinsic pulse shape characteristic of a particle is constant, and the overall pulse height is proportional to the total deposited energy.

\begin{figure}[H]
  \centering
  \includegraphics[,width=\textwidth]{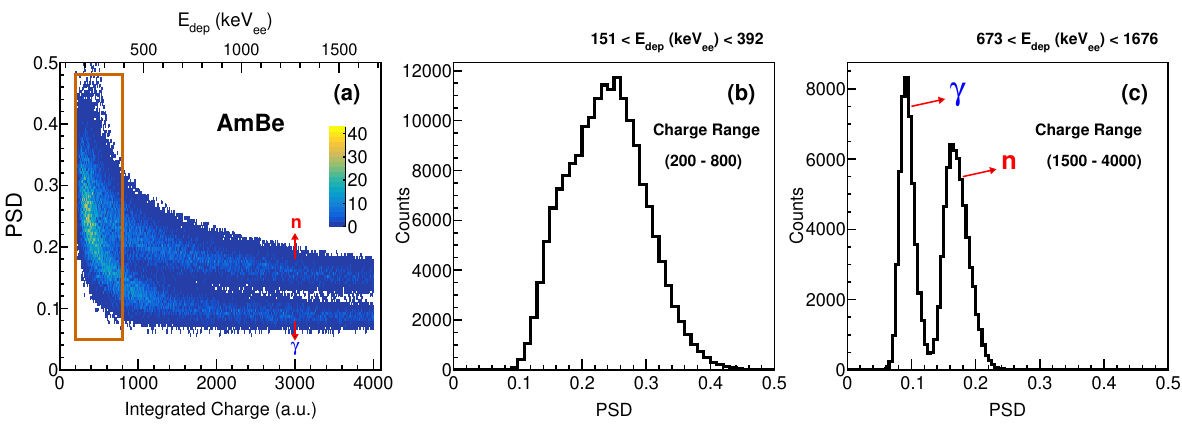}
  \caption{(a) PSD variable as a function of integrated charge, (b) projection of PSD variable for charge range between 200 to 800, and (c) projection of PSD variable for charge range between 1500 to 4000.}
  \label{PSDplots}
\end{figure}

The SG and LG time windows mentioned above are fixed through out the analysis. Figure~\ref{PSDplots} (a) shows the  PSD as a function of integrated charge measured from the AmBe neutron source. A clear separation between neutron and $\gamma$ events is observed above 1500 in the integrated charge range. The quantitative projection of PSD distributions are shown in Fig.~\ref{PSDplots} (b) and (c) for the lower charge and the higher charge ranges, respectively. It can be seen from Fig.~\ref{PSDplots} (b), that at lower charge values, between 200-800, the separation using PSD variable between neutron and $\gamma$ events worsens and the discrimination becomes difficult. Low signal to noise ratio, disparity in the Compton interaction, similar timing profiles of neutrons and $\gamma$s, resulting in effective failure of CCM in this region~\cite{Shen:2025vyt}. For the higher charge range (1500-4000), as shown in Fig.~\ref{PSDplots} (c), simple PSD based selection criteria can achieve a very good separation efficiency for discrimination of neutron and $\gamma$ events. 

\section{Machine Learning}
\label{MLapproach}
\subsection{Multivariate Analysis }
  As an alternate to PSD technique, machine learning algorithms are being explored in various experiments~\cite{Yun:2024fxv,Yu:2025lxz,BOREXINO:2023nsa} to distinguish different backgrounds. These methods do not directly rely on predefined charge integration windows, but can be trained to learn non-linear features from the raw waveforms. In this work, we will focus on the use of artiﬁcial neural networks (ANN), especially the MultiLayer Perceptron (MLP)~\cite{Bourlard1994} using a Bayesian inference, and a Support Vector Machine (SVM)~\cite{svm} algorithms for the separation of neutron and $\gamma$ events.

The MLP Bayesian Neural Network (MLPBNN) approach combines the structure of a traditional neural network with probabilistic modeling. In this framework, weights and biases are treated as probability distributions rather than fixed values. The network consists of one input layer, one or more hidden layers and a output layer. Bayesian inference is used to estimate these distributions, allowing the model to capture uncertainty in predictions. This approach improves generalization, offers uncertainty estimates, and reduces overfitting, especially in small or noisy datasets. MLPBNN is an efficient technique to tag the nonlinear correlation with associated uncertainties. This algorithm has been proven to be efficient for various practical application such as pattern recognition in image analysis~\cite{VEHTARI20001183}.
In parallel, SVM algorithm which is also a supervised ML algorithm, used for classification and regression tasks. It works by finding the optimal hyperplane that best separates data points of different classes in a higher-dimensional space. The key idea is to maximize the margin between the closest data points (support vectors) and the hyperplane. SVMs can also use kernel functions to handle non-linear data by mapping it into a higher-dimensional space where a linear separator is possible. SVMs are known for their accuracy and effectiveness in high-dimensional spaces.

While both methods have their advantages, MLPBNN needs fine-tuning for the involved hidden layers. Training MLPBNNs often takes more iterations to converge compared to traditional MLPs, particularly when using complex approximate inference techniques. Bayesian methods introduce additional hyperparameters (e.g., prior distributions, variational parameters) that require careful tuning, which can be non-trivial. On other hand, the SVM can be at many times slow to train and predict with large datasets, due to their complexity and memory requirements. The performance of SVMs heavily depends on the choice of kernel and parameters. They also do not provide direct estimation of probabilities for the predictions.
For our studies we are using the TMVA toolkit~\cite{hoecker2009tmva}, where the MLPBNN and SVM algorithms are readily available for the use. The advantage of using TMVA is the ease of integration of the packages to the ROOT based software from CERN~\cite{root}, which enables the easy data handling, computation and plotting the results. 
  
\subsection{Discrimination Variables}
To effectively discriminate neutron and $\gamma$ events from the AmBe source, different discrimination variables are considered as an alternative to the PSD technique. Fractions of integrated charges from different regions in time and cumulants~($C_i$) are calculated from the waveform for neutron and $\gamma$ as shown in Fig.~\ref{Waves}(b). These variables are used as input for ML algorithms. We have considered the ratio of the integrated charge in the time window of 48 ns to 68 ns to the total charge in the time window of 46 ns to 250 ns. This ratio is denoted as head charge ($R_{head}$~\cite{Lee:2024lvr}) variable and is given in Eq.(\ref{eq:rhead}).
\begin{align}
  R_{\text{head}} &= \frac{\sum_{j=48\,\text{ns}}^{68\,\text{ns}} Q_j}{\sum_{j=46\,\text{ns}}^{250\,\text{ns}} Q_j} \label{eq:rhead}
\end{align}

Similarly the charge weighted mean time in the time window of 78 ns to 108 ns is defined as $\langle t \rangle_{Q}$ variable, given in Eq.(\ref{eq:qmt}).
\begin{align}
  \langle t \rangle_{Q} &= \frac{\sum_{j=78\,\text{ns}}^{108\,\text{ns}} Q_j \cdot j}{\sum_{j=78\,\text{ns}}^{108\,\text{ns}} Q_j} \label{eq:qmt}
\end{align}

Since the differences in the waveforms of neutron and $\gamma$ mainly occur in the tail region, we have introduced a variable which captures the essence of tail part of the distribution. To evaluate this variable, we compute the cumulants of the waveform distribution up to fourth order.  The first several moments are called mean, variance, skewness, and kurtosis. The moments of the distributions are related to the cumulants as:~Mean ($\mu$) = $ C_1$;~Variance($\sigma^2$) = $ C_2$;~Skewness($S$) = $ C_3/C_2^{3/2}$;and Kurtosis($\kappa$) = $C_4/C_2^2$, where $C_i$ is related to the cumulants or the standardized $i^{th}$ moment about the mean. The ratio of the fourth-to second-order cumulants is the non-Gaussian fluctuations in the distribution, which is defined as: 
\begin{align}
  \kappa\sigma^2 &= \frac{C_4}{C_2} \label{eq:ksig2}  
\end{align}

The kurtosis by definition is a measure of relative tailedness of a distribution, which is a widely used parameter to study fluctuations in the hadron to quark gluon plasma phase transition by looking at the multiplicity distribution of produced particles in high energy heavy ion collisions~\cite{STAR:2010mib,Garg:2013ata}. When multiplied by the variance, it enhances the magnitude of the fluctuations arising from the differences in the tail part of the distribution. As neutrons have a comparatively longer tail part, thus, $\kappa\sigma^2$--given by Eq.(\ref{eq:ksig2})-- is considered as a suitable input variable for the discrimination using ML algorithms.
\begin{figure}[H]
  \centering
  \includegraphics[,width=\textwidth]{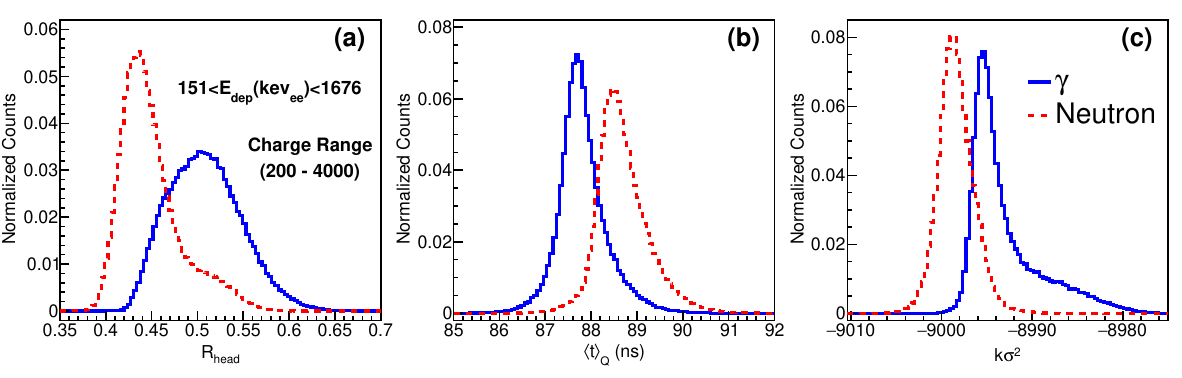}
  \caption{Input variables used for the ML algorithms are (a) $R_{head}$, (b) $\langle t \rangle_{Q}$ (in ns), and (c) $\kappa\sigma^2$. The distributions are shown for neutron (red dashed) and $\gamma$ (blue) events.}
  \label{Variables}
\end{figure}
Figure~\ref{Variables} (a), (b), and (c) shows the distribution of $R_{head}$, $\langle t \rangle_{Q}$(in ns), and $\kappa\sigma^2$ for neutron and $\gamma$ events, respectively. A reasonable degree of separation among different variables is observed for neutron and $\gamma$ events and hence can be used for training purposes of the ML algorithms.
\subsection{Training of dataset}
\begin{figure}[H]
  \centering
  \includegraphics[,width=\textwidth]{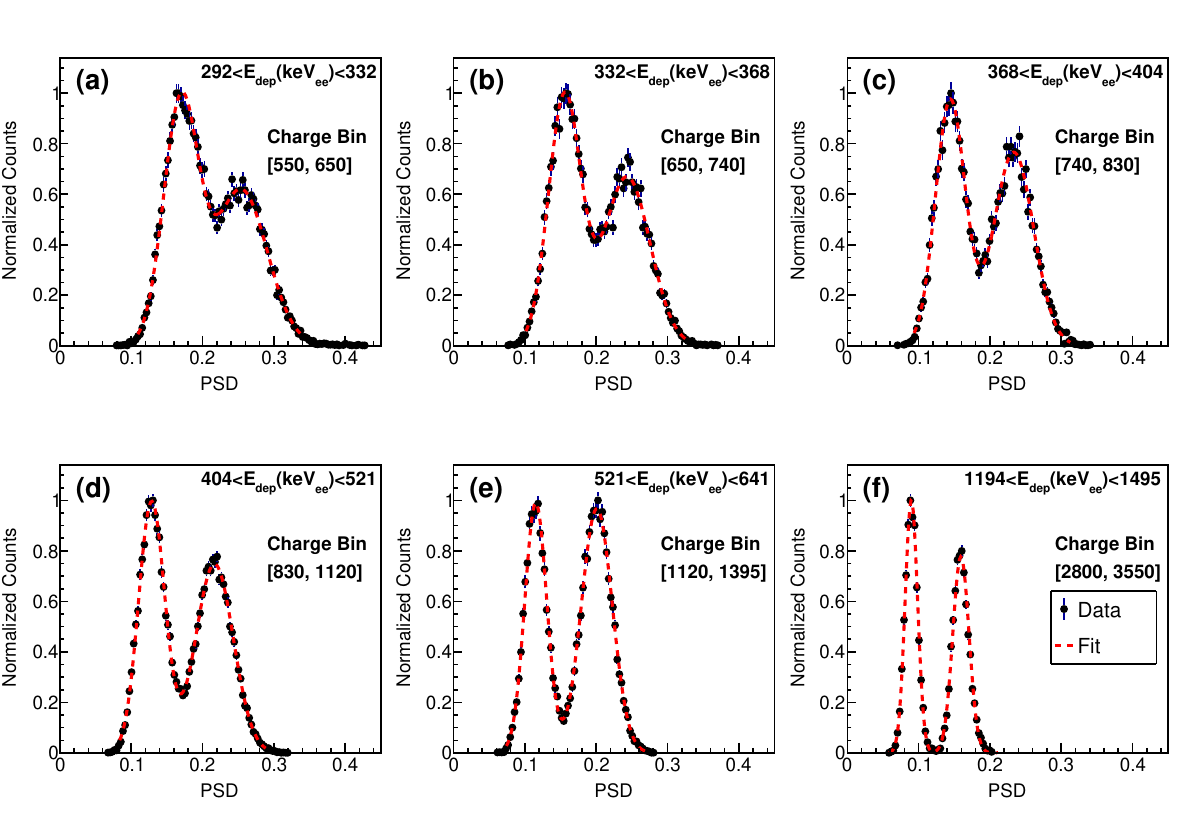}
  \caption{PSD distribution from AmBe source in different integrated charge ranges.}
  \label{PSDbins}
\end{figure}

\begin{figure}[H]
  \centering
  \includegraphics[,width=0.8\textwidth]{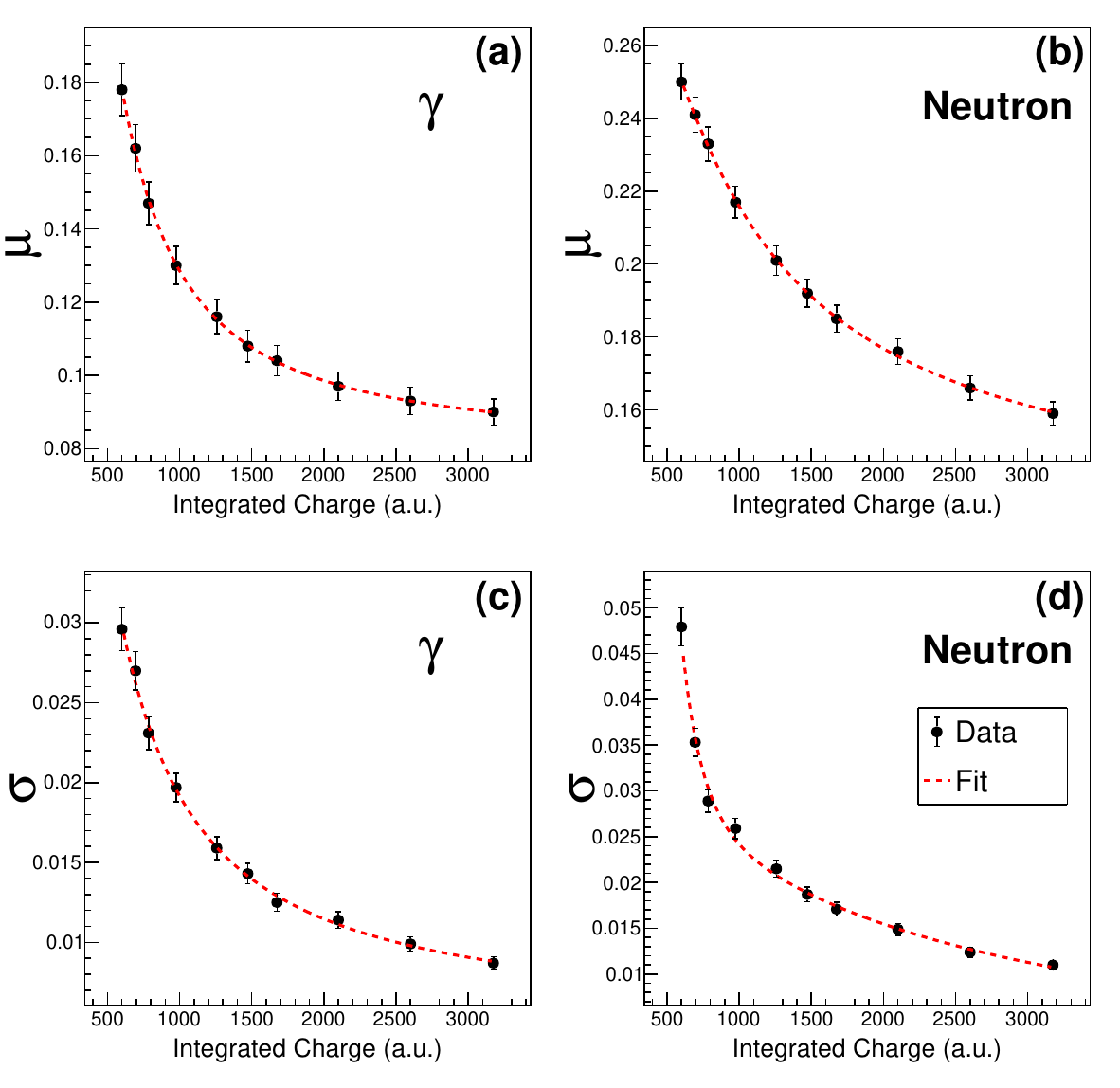}
  \caption{Variation of $\mu$ and $\sigma$ for the PSD distributions as a function of integrated charge for neutron and $\gamma$. The data are fitted with a polynomial function for the parameterization.} 
  \label{Params}
\end{figure}

For the supervised ML algorithms, the training of models require an input dataset with labeled signal and background events. To construct training dataset, a data driven PSD based model is considered for initial labeling. Figure~\ref{PSDbins} (a) to (f) shows the projection of PSD distribution in few selected different integrated charge ranges. The mean ($\mu$) and the standard deviation ($\sigma$) for neutron and $\gamma$ events, as a function of charge, are obtained for each charge range by fitting the PSD projection with a double Gaussian function. Figure~\ref{Params} (a)-(d) shows the variation of the $\mu_{n,\gamma}$ and the $\sigma_{n,\gamma}$ of neutron and $\gamma$ events as a function of integrated charge from 200 to 4000. The variation is parameterized by fitting the data with a third-order polynomial function and the interpolation of this function is used to obtain $\mu$ and $\sigma$ at lower charge region where the neutron and $\gamma$ peaks cannot be fitted explicitly.
Since the $\mu$ and $\sigma$ values show a variation as a function of integrated charge, we obtain a normalized variable, namely ``$\mathrm{kSigma}$'', for the selection of neutron and $\gamma$ input events for training of ML algorithms. The ``$\mathrm{kSigma}$'' variable is calculated using Eq.(\ref{eq:nsig}), where the $k$ represent the number of standard deviation around the mean ($\mu$) of PSD distribution which is adapted dynamically across different charge regions for the selection of training events.

\begin{align}
  \label{eq:nsig}
  (\mathrm{kSigma})_n &= \frac{PSD - \mu_n}{\sigma_n} , \qquad
  (\mathrm{kSigma})_\gamma = \frac{PSD - \mu_\gamma}{\sigma_\gamma}.
\end{align}
\begin{figure}[H]
  \centering
  \includegraphics[,width=\textwidth]{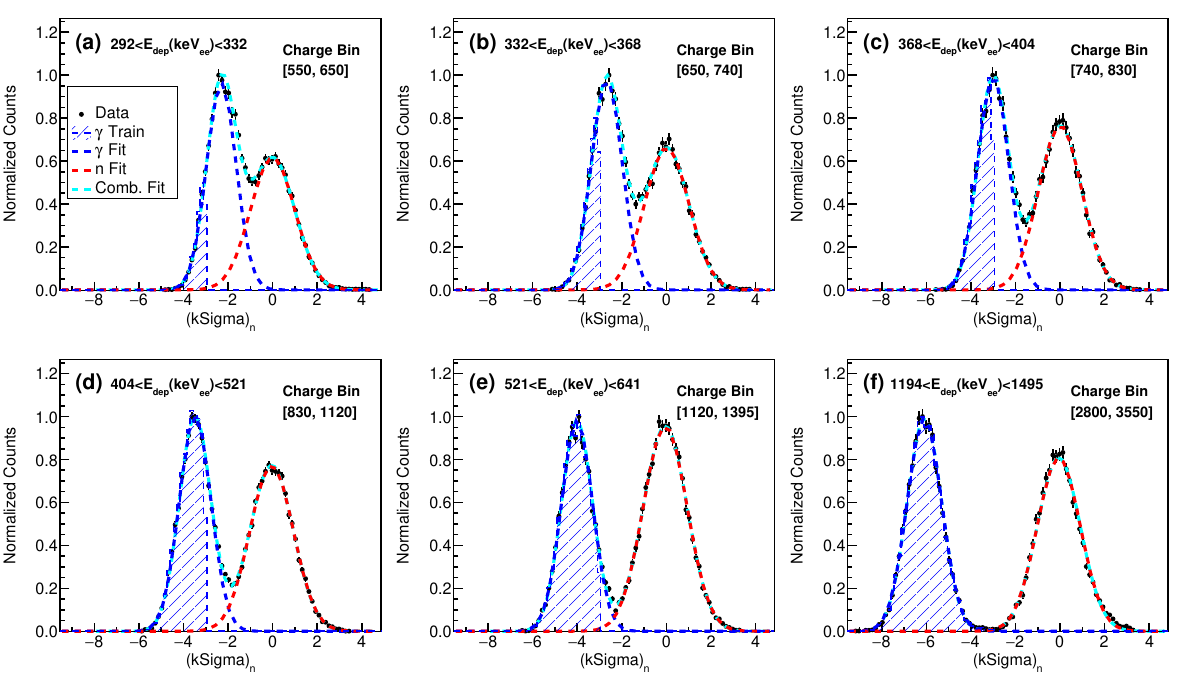}
  \caption{$\mathrm{(kSigma)_{n}}$ distribution is shown in different integrated charge bins. The combined fit (in teal) as well as $\gamma$ and neutron fit is shown in blue and red Gaussians, respectively. The shaded blue region represents the $\gamma$-events labeled for training.}
  \label{GamSelInbins}
\end{figure}
\begin{figure}[H]
  \centering
  \includegraphics[,width=\textwidth]{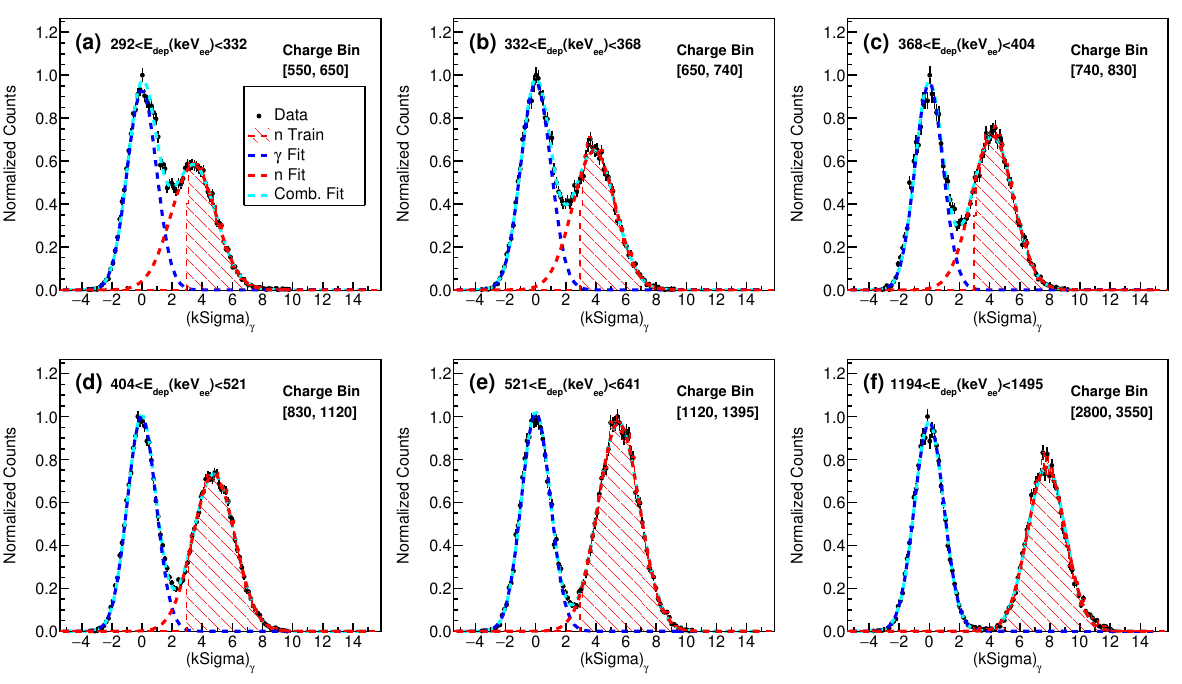}
  \caption{$\mathrm{(kSigma)_{\gamma}}$ distribution is shown in different integrated charge bins. The combined fit (in teal) as well as $\gamma$ and neutron fit is shown in blue and red Gaussians, respectively. The shaded red region represents the neutron-events labeled for training.} 
  \label{NeutSelInbins}
\end{figure}

The training dataset contains neutron and $\gamma$ events selected on the basis of $\mathrm{(kSigma)_{n,\gamma}}$ parameter. Figure~\ref{GamSelInbins} shows the $\mathrm{(kSigma)_{n}}$ distribution in different integrated charge bins. Also shown, are the combined Gaussian fit of the distributions along with the individual components representing the neutron and $\gamma$ event fractions. The events satisfying condition $\mathrm{(kSigma)_{n}}$ < -3.0, shown in blue shaded region, represents the selected true label for training of $\gamma$ events.
  Similarly, Fig.~\ref{NeutSelInbins} shows the $\mathrm{(kSigma)_{\gamma}}$ distribution in various integrated charge bins along with the Gaussian fits. The red shaded region in the Fig.~\ref{NeutSelInbins} (a)-(f) shows the selection of neutron events for the true label with the selection criteria of $\mathrm{(kSigma)_{\gamma}}$ > 3.0. With this selection criteria on $\mathrm{(kSigma)_{n}}$ and $\mathrm{(kSigma)_{\gamma}}$ for $\gamma$ and neutron true labels for training of ML algorithms, the uncertainty in the selection of true labels is reduced to less than $0.1\%$ in the studied integrated charge range.

For each ML algorithm, a training dataset is prepared by randomly selecting 20,000 events each for neutron and $\gamma$ based on the above selection. Another dataset, containing 100,000 neutron and $\gamma$ events are used for the testing purpose of the ML algorithms. To identify the optimal configuration, for both MLPBNN and SVM algorithms, optimization is done by tuning the set of input parameter dependencies and stability of the algorithms on the variables of interest.

For MLPBNN, the convergence of the algorithm is tested with different numbers of hidden layers.The first three hidden layers consist of N+4, N+4, and N+2 nodes where N represents the number of variables. Apart from the three input variable nodes, MLPBNN uses a bias node. A bias node provides a constant input to the nodes in the next layer and have their own weights associated with each connection which is also learned during training. The activation function in our case is a non linear sigmoid function which is represented by $\mathrm{tanh~(x)}$. This function introduces a nonlinearity into the network, allowing it to learn complex patterns in the data. In order to reduce the number of iterations to optimize the computation time an alternative approach called the Broyden-Fletcher-Goldfarb-Shannon (BFGS) method is utilized while adapting the synapse weight. The combination of this particular architecture is found to be optimal in terms of performance and convergence of the network. The MLPBNN is trained for 200 epochs using the BFGS optimization method. The number of epochs are optimized by monitoring the performance of the algorithm and by comparing the outputs from training and testing datasets to check underfitting or overfitting of the results. Additionally, the degree of correlation between the variables used for training was optimized to minimize the potential bias.
The MLPBNN provides a method for estimating the relative importance of each input variable through a variable ranking technique. This method utilizes the sum of the squared weights of the connections between the input neuron corresponding to variable \( i \) and all neurons in the first hidden layer. The importance \( I_{i} \) of the input variable \( i \) is defined as,

\begin{equation}
I_{i} = x_{i}^2 \sum_{j=1}^{n_h} \left(w_{ij}^{(1)}\right)^2, \quad i = 1, \dots, n_{\text{var}}
\label{eq:importance}
\end{equation}
where, \( x_i \) is the sample mean of the input variable \( i \), \( w_{ij}^{(1)} \) is the weight connecting input neuron \(i \) to hidden neuron \( j \) in the first hidden layer, \( n_h \) is the number of neurons in the first hidden layer, and \( n_{\text{var}} \) is the total number of input variables. The importance of the three input variables are presented in the table~\ref{varImp}. It can be seen that the input parameter $\kappa\sigma^2$ provides better ranking in terms of separability than other variables.

\begin{table}[H]
  \centering
  \caption{Ranking of input variables for MLPBNN.}
  \label{varImp}
  \begin{tabular}{lcc}
    \toprule
    \textbf{Rank} & \textbf{Variable} & \textbf{Importance (in\%)} \\
    \midrule
    1    & $\kappa\sigma^{2}$             & 61.03 \\
    2    & $\langle t \rangle_{Q}$   & 47.28 \\
    3    & $R_{head}$                 & 43.76 \\
    \bottomrule
  \end{tabular}
\end{table}

In parallel, the SVM algorithm is tested with Radial Basis Function (RBF) kernel mode. The RBF kernel is usually used when the data are non linear in nature. Apart from the calculation of the squared distances between points, it has the $\Gamma$ parameter which controls how well model fits the data. For SVM algorithm, the tolerance and $\Gamma$ values are tuned to minimize the false positive rate which is given by Type-I error. The RBF mode used in our training dataset has a $\Gamma$ value of 0.25 and with a tolerance of 0.001. Though higher $\Gamma$ (=1) value yields marginal gain ($< 0.25\%$) in area under the curve, but it introduces an overfitting risk. Thus we have chosen a value for which the decision boundary is smooth enough to classify efficiently while preserving the model from over-fitting.
\subsection{Response from MLPBNN and SVM}
\begin{figure}[H]
  \centering
  \includegraphics[,width=\textwidth]{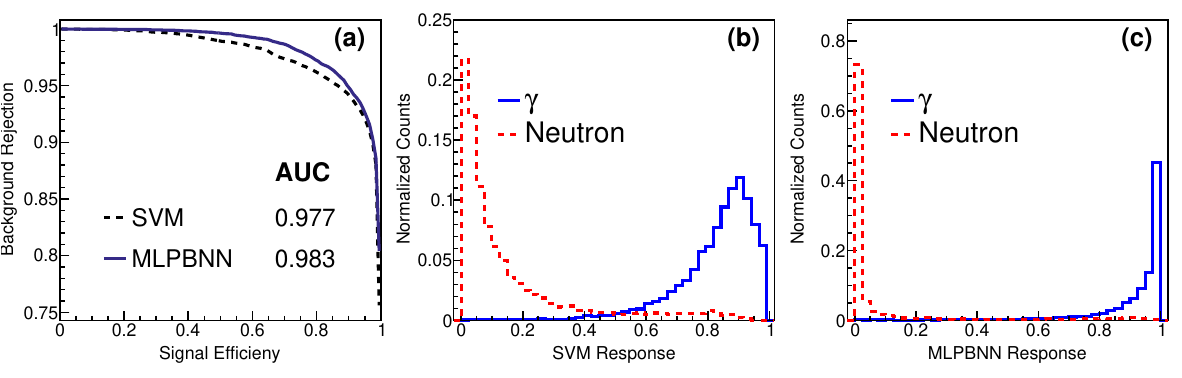}
  \caption{(a) Comparison of the ROC curves for MLPBNN and SVM algorithm. The classifier response of (b) SVM and (c) MLPBNN for neutron and $\gamma$ events after training of ML algorithms.}
  \label{Resp}
\end{figure}
Figure ~\ref{Resp} (a) shows the comparison of the Receiver Operating Characteristic (ROC) curve for the MLPBNN and SVM algorithm. The background rejection efficiency or False Positive Rate (FPR) is represented as a function of the signal selection efficiency or the True Positive Rate (TPR). From the ROC curve, MLPBNN is found to achieve better background rejection for the same signal efficiency as compared to SVM.
Also shown, Fig.~\ref{Resp} (b) and (c), are the classifier responses for SVM and MLPBNN methods, respectively. A reasonable level of discrimination between the neutron and $\gamma$ events is achieved from the ML algorithms with our variables and optimized parameters.
\begin{table}[htbp]
  \centering
  \caption{Optimal cut for selection of neutron and $\gamma$ events and significance of the ML algorithms.}
  \label{MLcuts}
  \begin{tabular}{lcc}
    \toprule
    \textbf{ML Algorithm} & \textbf{Optimal Cut} & \textbf{Significance \(\left(\frac{S}{\sqrt{S+B}}\right)\)} \\
    \midrule
    MLPBNN &  0.4495 & 29.81 \\
    SVM    &  0.4641 & 29.80 \\
    \bottomrule
  \end{tabular}
\end{table}
The optimal cuts, as shown in Table~\ref{MLcuts}, for both models are used in the analysis to discriminate between neutron and $\gamma$ events. The significance of around 30 is achieved in both the algorithms with 1000 neutron and 1000 $\gamma$ events.

\begin{figure}[H]
  \centering
  \includegraphics[,width=0.7\textwidth]{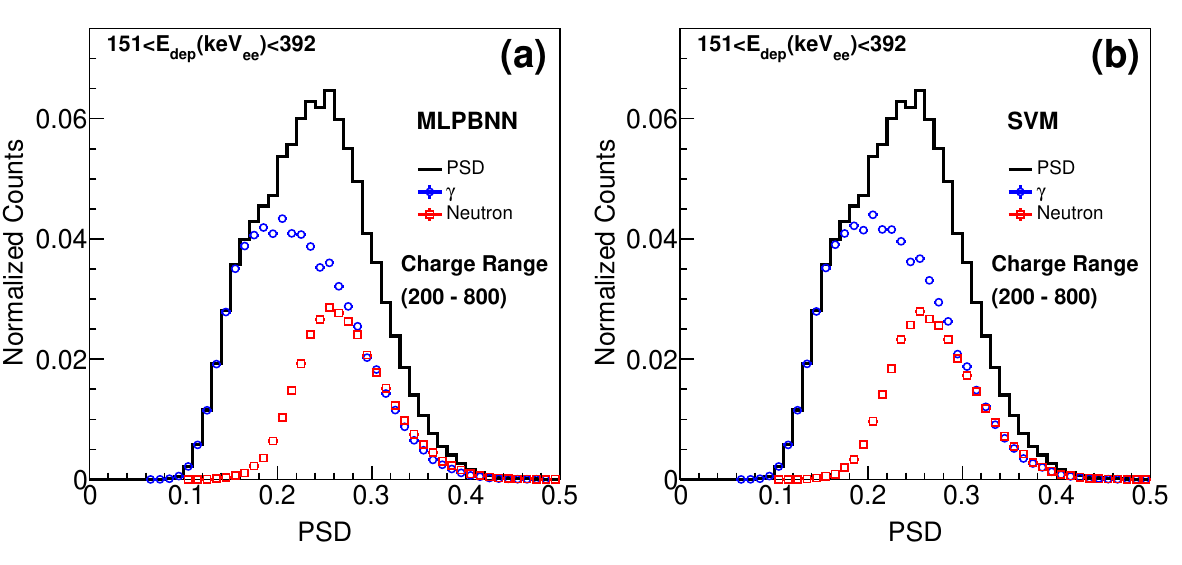}
  \caption{Comparison of PSD projection of neutron and $\gamma$ events discriminated using ML score and CCM in the charge range (200-800) for (a) MLPBNN and (b) SVM algorithm.}
  \label{PSD1dLE}
\end{figure}

\begin{figure}[H]
  \centering
  \includegraphics[,width=0.7\textwidth]{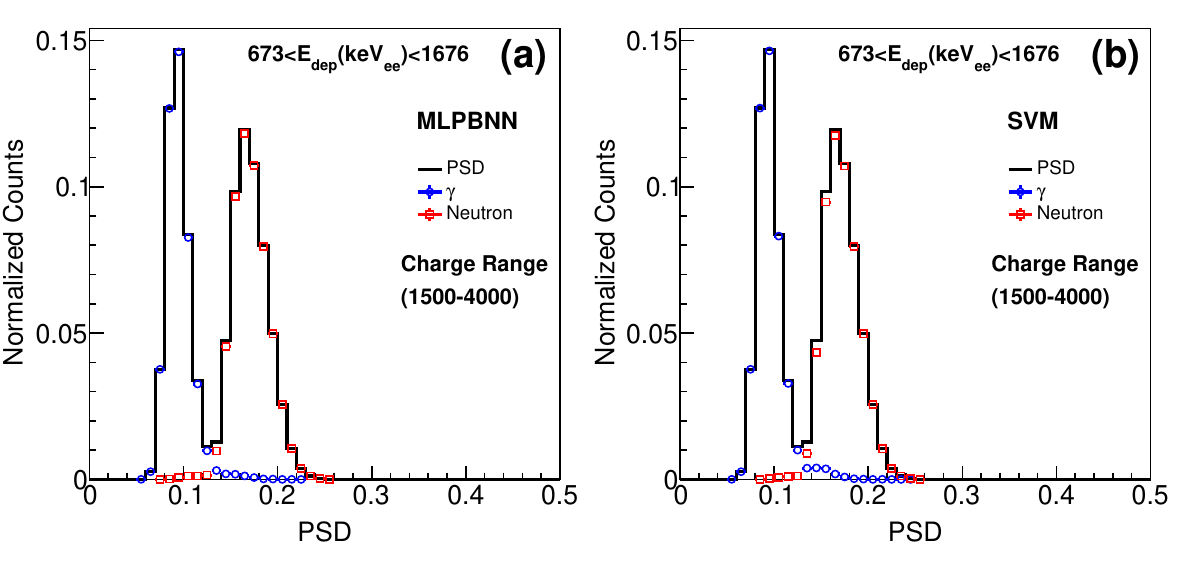}
  \caption{Comparison of PSD projection of neutron and $\gamma$ events discriminated using ML score and CCM in the charge range (1500-4000) for (a) MLPBNN and (b) SVM algorithm.}
  \label{PSD1dHE}
\end{figure}

The application of the obtained weights, from the training of the MLPBNN and SVM algorithm, on the remaining dataset variables $R_{head}$, $\langle t \rangle_{Q}$ and $\kappa\sigma^2$ are calculated for each waveform on event by event basis. The discrimination of the neutron and $\gamma$ events, using the values from table~\ref{MLcuts}, is achieved from the ML score obtained for both the algorithms.
Figure~\ref{PSD1dLE} (a) and (b) show the PSD distribution of events obtained with the discrimination by the application of ML-score from MLPBNN and SVM algorithm for the charge range 200 to 800, respectively. The solid histogram show the PSD variable calculated from the CCM where the separation between neutron and $\gamma$ events is not distinguishable. The red open square show the discriminated neutron events and the blue open circle show the $\gamma$ events for MLPBNN ( in Fig~\ref{PSD1dLE} (a) ) and for SVM ( in Fig~\ref{PSD1dLE} (b) ) algorithm using the ML score.
Figure~\ref{PSD1dHE} (a) and (b), show the discriminated neutron and $\gamma$ events for the charge range 1500 to 4000, between the conventional CCM and ML algorithms. The discrimination of neutron and $\gamma$ events in both CCM and ML algorithms agrees well, demonstrating the accuracy of the results obtained from the ML score in the higher charge region.

For the benchmarking of the results obtained from ML algorithms in lower charge range, we discriminate the neutron and $\gamma$ events based on Time-of-flight (ToF) technique. In ToF, neutron and $\gamma$ events can be explicitly separated in the lower charge ranges as opposed to conventional CCM. This would allow us to tag the pure neutron and $\gamma$ events from ToF which would help us to estimate the performance and errors for the predictions of neutron and $\gamma$ events from ML algorithms. 

\section{Time of Flight Experiment}
\begin{figure}[H]
  \centering
  \includegraphics[,width=\textwidth]{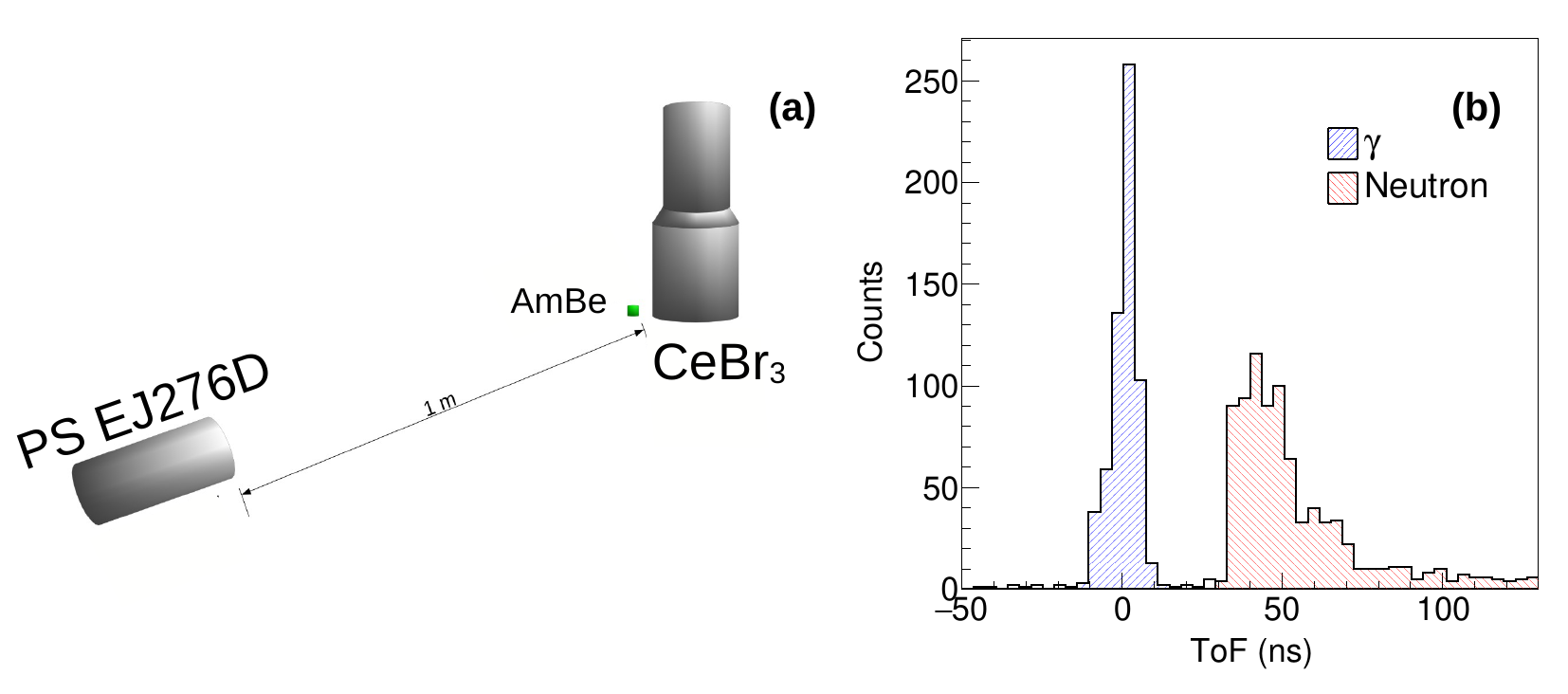}
  \caption{(a) Experimental schematic of the ToF experimental setup and (b) ToF distribution obtained by the timing difference between the start and stop detector.}
  \label{ToF}
\end{figure}

For the ToF experiment, a 3-inch $\mathrm{CeBr_3}$ crystal coupled to a Hamamatsu 10-stage PMT was used as start time detector. Since the detector has excellent energy resolution, $\sim$4$\%$ at 0.662 MeV~\cite{Dey:2022fza}, was used to trigger on the 4.4 MeV $\gamma$ events from the AmBe neutron source. The same PS detector was used as the stop time detector. Both the detectors were positioned at 1 m away from each other. The AmBe source was kept near the $\mathrm{CeBr_3}$ start time detector. For Figure~\ref{ToF} (a) shows the schematic diagram of the ToF experimental setup. The kinetic energy of neutrons \( \mathrm{E_{n}} \) is given in Eq.~\ref{eq:neutron_energy}.

\begin{equation}
  \label{eq:neutron_energy}
  \mathrm{E_{n} = \frac{1}{2} m_n v^2 = \frac{1}{2} m_n \left( \frac{L}{ToF} \right)^2} ,
\end{equation}
where, \(\mathrm{ m_n}\) is the mass of the neutron, \( \mathrm{v}\) is the velocity of the neutron, \( \mathrm{L} \) is the flight path length, and \( \mathrm{ToF} \) is the time-of-flight. For 1 meter flight path, time-of-flight for neutron and $\gamma$ events is given by,

\begin{equation}
  \label{eq:tof_neutron}
  \text{ToF}_n = \frac{72.3 }{\sqrt{E_n}},\quad \text{ToF}_\gamma = \frac{1}{c}.
\end{equation}

\begin{figure}[H]
  \centering
  \includegraphics[,width=0.6\textwidth]{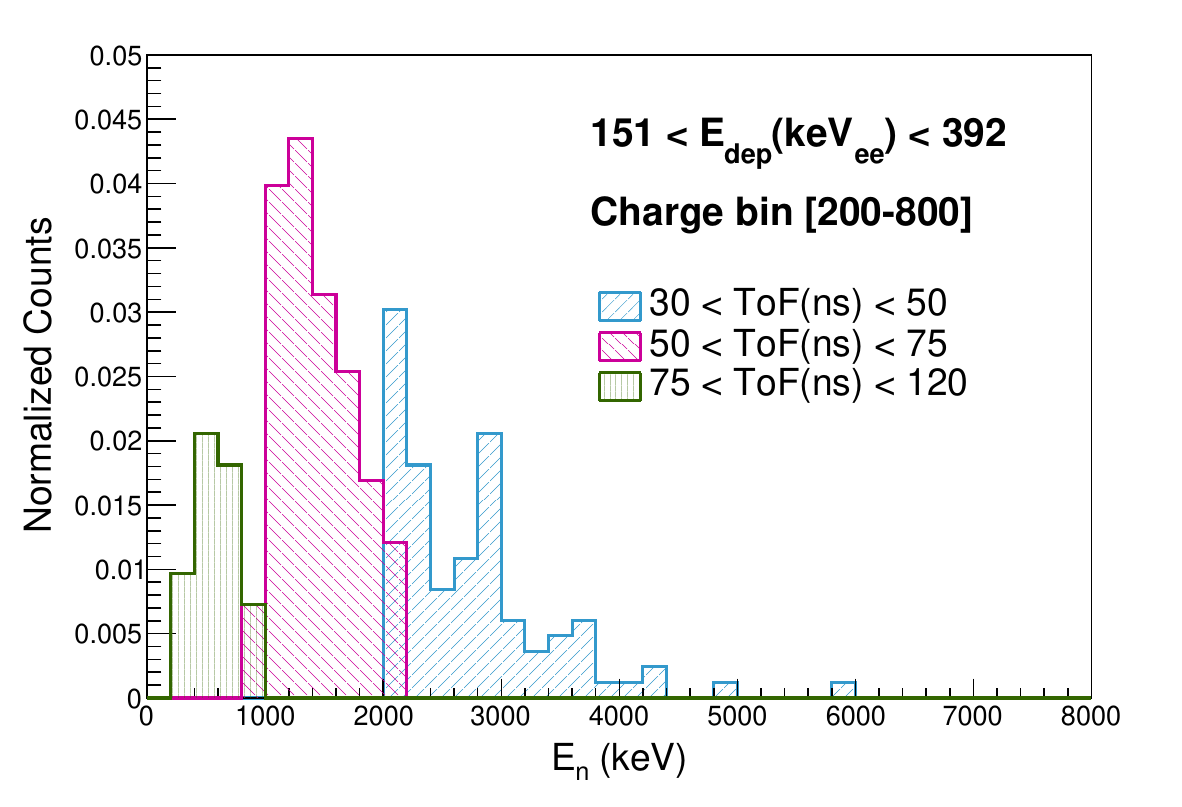}
  \caption{Neutron energy distribution, calculated from ToF, for different ToF bins having an integrated charge in 200 to 800.} 
  \label{NeutInTimebins}
\end{figure}

The trigger scheme used for the ToF measurement in the waveform digitizer was configured such that a start signal timestamp was generated when the $\mathrm{CeBr_3}$ detector registers a 4.4 MeV $\gamma$ event. The stop signal timestamp was recorded when either a neutron or a $\gamma$ event was triggered in the PS detector. To reduce the rate of accidental background events, a correlated time window of 200 ns ($\mathrm{E_n \sim 130 }$ keV) interval between start and stop time was used for the data acquisition. Due to the relatively small active area of the PS detector, the overall trigger acceptance of the setup was relatively low and the data presented for the ToF analysis consists of 3 days.~The raw waveforms from both the detectors were recorded and the PS detector data was analyzed in the same manner as for the singles events. The ToF variable was calculated by taking the recorded start and stop timestamp generated from $\mathrm{CeBr_3}$ and PS detector, respectively.
In the charge range 200 to 4000, the ToF distribution shows a clear separation between the extracted neutron and $\gamma$ events, as shown in Fig.~\ref{ToF}(b). The events with $\mathrm{ToF > 30}$ ns are tagged as neutron events~(red hatched region), while events with $\mathrm{ToF < 25}$ ns are tagged as $\gamma$ events~(blue hatched region). For the integrated charge range 200 to 800, a total of 650 $\gamma$ and neutron events are filtered. For the same charge range, Fig.~\ref{NeutInTimebins} shows the neutron energy distribution, calculated from Eq.~\ref{eq:tof_neutron}. The energy deposition by a recoiling proton is affected by the finite detector acceptances, charge collection inefficiencies, and quenching effects arising from non-linear energy deposition. Consequently, a fraction of neutron events ($\sim 36\% $) are found to have proton recoil energy deposition in the lower $\mathrm{E_{dep}}$ range of 150 to 400 $\mathrm{keV_{ee}}$ where the conventional PSD technique fails to discriminate neutrons and $\gamma$-rays.
The PSD parameter is then calculated for these tagged events, in the same manner as done for singles events, and are compared with the PSD distribution obtained from the discrimination of neutron and $\gamma$ events using the ML score.
\begin{figure}[H]
  \centering
  \includegraphics[,width=\textwidth]{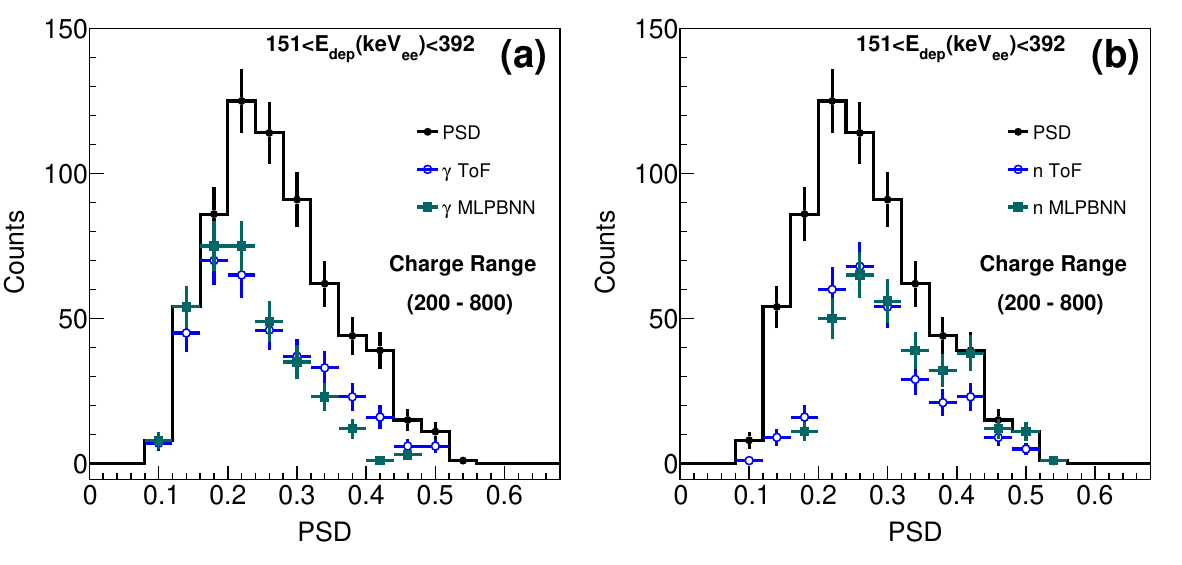}
  \caption{(a) Comparison of PSD distribution between CCM (black), $\gamma$ tagged events (blue circle) from ToF and from the MLPBNN score (green square). (b) Comparison of PSD distribution between CCM (black), neutron (n) tagged events (blue circle) from ToF and from the MLPBNN score (green square).}
  \label{mlpResp}
\end{figure}

\begin{figure}[H]
  \centering
  \includegraphics[,width=\textwidth]{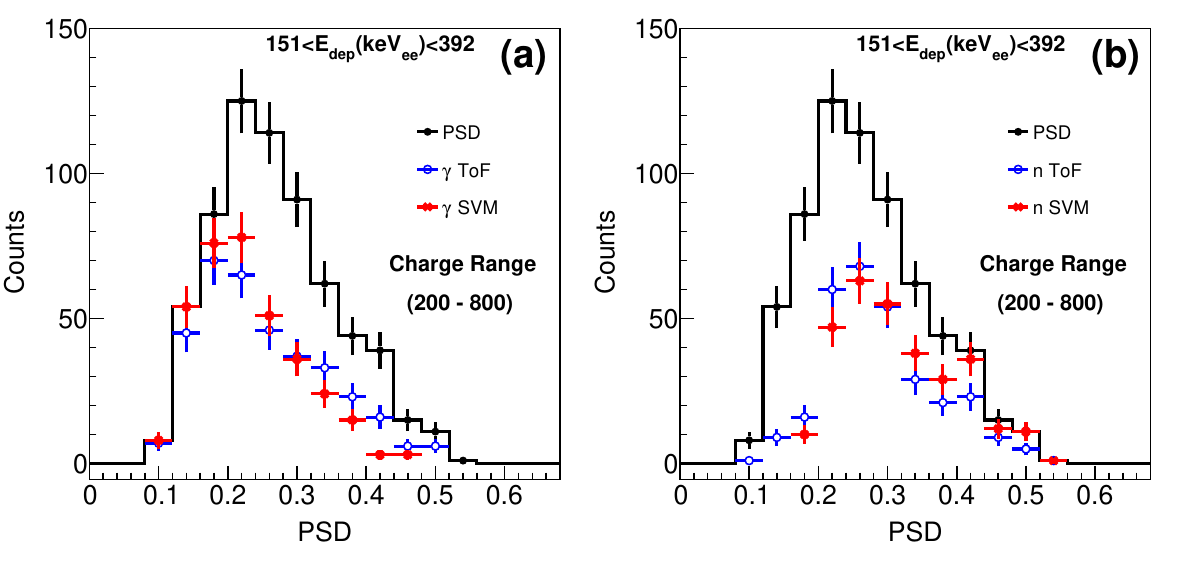}
  \caption{(a) Comparison of PSD distribution between CCM (black), $\gamma$ tagged events (blue circle) from ToF and from the SVM score (red circle). (b) Comparison of PSD distribution between CCM (black), neutron (n) tagged events (blue circle) from ToF and from the SVM score (red circle).}
  \label{svmResp}
\end{figure}
Figure~\ref{mlpResp} (a) and (b) show the PSD variable calculated using the CCM (black solid histogram) for PS detector in charge range from 200 to 800. Also shown are the PSD distribution for neutron (Fig~\ref{mlpResp} (b)) and $\gamma$ (Fig~\ref{mlpResp} (a)) events discriminated using the MLPBNN (solid green square) and ToF (open blue circles) method, respectively. The agreement between the results obtained for the discrimination of neutron and $\gamma$ events from MLPBNN and ToF is very good within the statistical uncertainties. This demonstrates the robustness of the ML algorithm for the efficient discrimination of neutron and $\gamma$ events where the CCM fails. Similarly, Fig~\ref{svmResp} (a) and (b) show, the PSD variable calculated from CCM (black solid histogram) and comparison of discriminated neutron and $\gamma$ events obtained from SVM algorithm and ToF method. Qualitatively, both the ML algorithms discriminate neutron and $\gamma$ events reasonably well in the studied lower energy regime.
\section{Discussion}
\begin{figure}[H]
  \centering
  \includegraphics[,width=\textwidth]{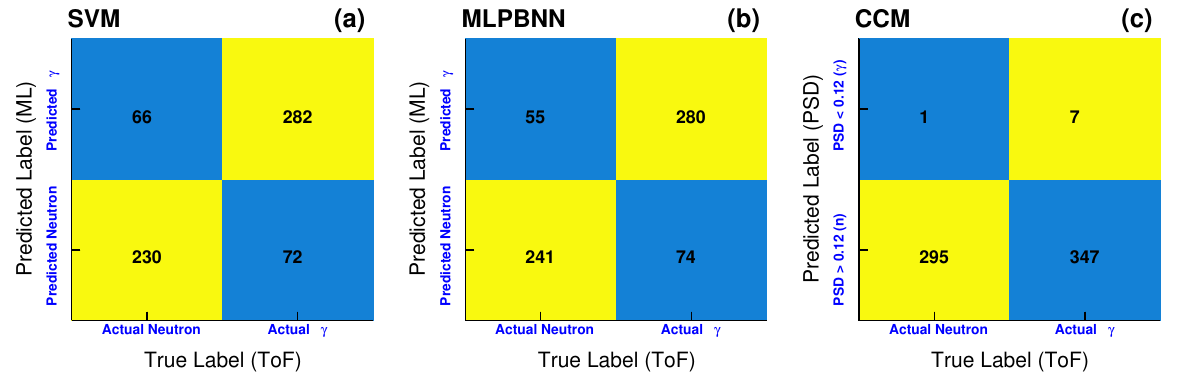}
  \caption{Confusion Matrix for (a) SVM, (b) MLPBNN, and (c) CCM is shown for the integrated charge range 200-800. The yellow quadrants represent the correctly classified events (True Positives and True Negatives), and the blue quadrants report the misclassified events (False positives and False Negatives).  }
  \label{ConfMat}
\end{figure}
To quantify the robustness of our ML algorithms, we calculate the confusion matrix as shown in Fig.~\ref{ConfMat} (a), (b), and (c), which represents the discrimination performance of the SVM, MLPBNN, and CCM, respectively. The true labels in the x-axis are obtained from the ToF technique and the y-axis represents the predicted labels from ML algorithms or PSD. The PSD-based selection of neutron and $\gamma$ events in the integrated charge range 200-800, is motivated from the well separated bands in the higher charge range. The true positives (TP) define the actual $\gamma$ events correctly classified as $\gamma$, and the false positives (FP) correspond to the neutron events incorrectly predicted as $\gamma$ events. Similarly, the true negatives (TN) defines the actual neutron events which are correctly classified as neutron events and false negative (FN) corresponds to the $\gamma$ events incorrectly classified as neutron events. From this formulation, we compute some of the key measures for the MLPBNN, SVM algorithms, and CCM which are given in the Table.~\ref{Specs}. The accuracy, defined as the ratio of sum of TP and TN to the total number of events used for the classification. It gives the overall performance of the algorithm in discrimination of the neutron and $\gamma$ events. Precision is defined as the ratio of TP and the sum of TP and FP, and represents the quality of the models positive predictions. Similarly recall is defined as the ratio of TP and the sum of TP and FN, measuring the probability of predicting positives. Specificity is another important parameter which evaluates the classification of the models particularly in binary classification as in our case. It measures the ability of a model to correctly identify the negative instances and is calculated by taking the ratio of TN to the sum of TN and FP.
\begin{table}[htbp]
  \centering
  \caption{Metrics for the ML algorithms used for classification.}
  \label{Specs}
  \resizebox{\textwidth}{!}{%
    \begin{tabular}{lccccccccc}
      \toprule
      \textbf{ML Algorithm} & \textbf{Accuracy} & \textbf{Precision} & \textbf{Recall} & \textbf{Specificity} & \textbf{F1 Score} & \textbf{Type-I} & \textbf{Type-II} \\
      \midrule
      MLPBNN  & 0.80 & 0.84 & 0.79 & 0.81 & 0.81 & 0.19 & 0.21 \\ 
      SVM     & 0.79 & 0.81 & 0.80 & 0.78 & 0.80 & 0.22 & 0.20  \\
      CCM     & 0.46 & 0.88 & 0.02 & 0.99 & 0.04 & 0.01 & 0.98\\
      \bottomrule
    \end{tabular}
  }
\end{table}
Lastly, the F1-score provides a measure of performance of the model particularly when there is an imbalance of events in the datasets. The F1-score is represented by $\mathrm{ 2 \times Precision \times Recall / (Precision+Recall)}$. Apart from these performance metric to evaluate our ML algorithms, we also define the type-I and type-II errors. Type-I error is calculated as ratio of FP to the sum of FP and TN and effects the precision of a model which measured accuracy of positive predictions. Similarly, type-II errors calculated from the ratio of FN to the sum of TP and FN denotes the impact on the recall of a model by signifying how well the model can identify all the actual predictions. In Fig.~\ref{ConfMat}, the yellow entries indicate the correctly classified entries, contributing to an overall accuracy of $80\%$ and $79\%$ for MLPBNN and SVM, respectively. Similarly, the blue entries representing falsely classified entries FP and FN, contribute to recall and precision, respectively. Overall, the key metrics for the ML algorithms used for the neutron and $\gamma$ discrimination are reasonable and shows better discrimination performance as compared to the CCM. This hints a promising future for  the use of such algorithms for the particle identification in lower energy deposition regions.
\section{Conclusion}
In this study, machine learning based algorithms are used on signals obtained from plastic scintillator detector with PSD capabilities. To enhance the discrimination of neutron and $\gamma$ events in the low energy regime where the traditional PSD technique has inferior performance, MLPBNN and SVM, are used to separate them using raw waveforms. The variables, $\mathrm{R_{head}}$, $\langle t \rangle_{Q}$, and $\mathrm{\kappa\sigma^{2}}$, calculated directly from the waveform, show promising improvements in the discrimination of neutron and $\gamma$ events at lower energies beyond 151 $\mathrm{keV_{ee}}$. Additionally, to cross-check the validity of the discrimination obtained from machine learning algorithms, a comparison of results for neutron and $\gamma$ events is done with the results obtained from a time-of-flight experiment. The time-of-flight experiment is particularly effective in the low energy regime, hence providing a platform to benchmark the ML algorithms.
The results from ML score and from ToF data shows an reasonable level of separation of neutron and $\gamma$ events in the low energy regime. Results obtained from the ML algorithms are consistent with those obtained from the ToF experiment, with an accuracy of 0.80 and 0.79 for MLPBNN and SVM, respectively. From the ROC curve as well as the sensitivity test of ML models, MLPBNN is found to have better discriminatory power in all charge regions.
For future studies, we aim to explore more advanced ML techniques such as Convolution Neural Network (CNN) and Deep Neural Network (DNN) to access the discrimination in even lower energy ranges with the improvements in lowering of the detector and electronic thresholds from the plastic scintillator detectors.


\section{Acknowledgments}
S.~P. thanks Department of Atomic Energy, India for the financial support.

\bibliographystyle{elsarticle-num}
\bibliography{refs_nim}

\end{document}